# Large spin signal and spin rectification in folded-bilayer graphene

Md. Anamul Hoque[1Ƭ], Zoltán Kovács-Krausz[2 Ƭ, 3], Bing Zhao[1], Prasanna Rout[1], Ivan Vera Marun[4,5], Szabolcs Csonka[2], Péter Makk[2,3*], Saroj P. Dash[1,6,7*]

[1]Department of Microtechnology and Nanoscience, Chalmers University of Technology, SE-41296, Göteborg, Sweden.
[2]Department of Physics, Institute of Physics, Budapest University of Technology and Economics, H-1111 Budapest, Hungary.
[3]MTA-BME Correlated van der Waals Structures Momentum Research Group, Műegyetem rkp. 3., H-1111 Budapest, Hungary.
[4]Department of Physics and Astronomy, University of Manchester, M13 9PL Manchester, United Kingdom.
[5]National Graphene Institute, University of Manchester, M13 9PL Manchester, United Kingdom.
[6]Wallenberg Initiative Materials Science for Sustainability, Department of Microtechnology and Nanoscience, Chalmers University of Technology,SE-41296, Göteborg, Sweden.
[7]Graphene Center, Chalmers University of Technology, SE-41296, Göteborg, Sweden.

## Abstract

Graphene is a promising platform for spin-based non-volatile memory, logic, and neuromorphic computing by combining long-distance spin transport with electrical tunability at room temperature. However, advancing beyond passive spin channels requires devices capable of generating large spin signals with efficient rectification capabilities, which are essential for active spintronic components. Here, we demonstrate a folded-bilayer graphene spin-valve device with giant non-local spin signals in the several mV range with pronounced spin-rectification effects. Efficient spin injection creates a significant spin accumulation of 20 meV and generates a spin diode effect with an asymmetry of over an order of magnitude between forward and reverse bias conditions. This spin-diode effect is expected to arise from nonlinear spin-charge interactions in the folded-bilayer graphene channel. These observed large spin signals and spin-diode effects in graphene systems offer a promising platform for developing active two-dimensional spintronic devices.

Ƭ **Equally contributed:** Md. Anamul Hoque, Zoltán Kovács-Krausz
***Corresponding authors:** makk.peter@ttk.bme.hu, saroj.dash@chalmers.se

## Introduction

Spintronic devices offer potential for next-generation computing by integrating non-volatile data storage with processing capabilities[1] for advanced spin-based logic[2–5], memory-in-computing, and neuromorphic computing architectures[6–10]. While substantial progress has been made in fundamental spin transport phenomena for information storage and non-volatile memory technologies[3,9,11,12], the field can now transition toward active and multifunctional spin-based logic and neuromorphic systems. These advanced applications demand devices capable of generating large-amplitude spin signals and spin amplification and rectification effects[6–10]. Graphene has emerged as a unique material for spin transport, offering room-temperature spin diffusion lengths exceeding 25 μm and lifetimes of over 10 ns, making it suitable for advanced spin-based devices[12–16]. Furthermore, wafer-scale demonstrations using chemical vapor deposition (CVD) graphene have shown spin communication over distances of over 30 μm[17–19], marking a significant progress toward multifunctional spin logic functionalities[4,20]. However, practical spintronic devices require two critical advances: first, achieving spin signal amplitudes in the millielectronvolt range, and second, implementing efficient spin amplification and rectification functionalities[21].

To address these active spin functionalities for cascading in spin logic applications[4,22], nonlinear interaction between spin and charge currents in the graphene channel should be maximized[23,24]. For example, the spin drift effect[25] exemplifies this nonlinear coupling, enabling detection of spin accumulation using nonmagnetic contacts and spin-to-charge conversion[26,27], the manipulation of spin currents using electric fields[28], spin signal amplification in bipolar devices[29,30] and directional control of spin currents[31]. Importantly, realizing strong spin-drift interactions requires graphene channels that combine a tunable and energy-sensitive conductivity, resilience to high electric fields, and the ability to create a substantial spin accumulation[32]. To achieve the performance target of a large spin signal amplitude with a large spin accumulation, it is essential to improve impedance matching between ferromagnetic contacts and graphene channel properties[13,32,33]. While considerable effort has focused on engineering tunnel barriers and their resistances to optimize impedance matching on graphene[23,24,34–39], the strategic tailoring of graphene channel properties to modulate resistance and enhance spin signal generation remains largely underexplored.

Here, we report a significant output spin signal with large rectification effects using a folded-bilayer graphene spin-valve device at room temperature. Such a folded-bilayer graphene channel provides near-ideal spin impedance matching conditions with ferromagnetic tunnel contacts, enabling efficient spin injection with a considerable spin accumulation ($\Delta\mu$) of ~ 20 meV in graphene with an estimated spin polarization of the ferromagnetic tunnel contact of ~24.5%. This pronounced spin-diode response demonstrates efficient nonlinear spin-signal processing with 2D spintronic devices.

## Results

### Observation of a large non-local spin signal:

We fabricated a spin-valve device using an exfoliated folded-bilayer graphene with ferromagnetic (FM) tunnel contacts ($TiO_2$/Co) on $SiO_2$/Si substrate, as shown in Figure 1a,b (see Methods and Supplementary Figure S1). The non-local (NL) measurement geometry (Figure 1b) was adopted for reliable detection of pure spin signal, where the injected current (*I*) and detected voltage ($V_{nl}$) circuits are separated with a graphene channel length L = 1.55 μm in Dev 1 (mainly used for data in the main manuscript). Figure 1c shows the optical microscope pictures of the folded graphene on $SiO_2$/Si substrate. There are approximately 2-3 folds of graphene bilayers in the channel, as estimated by comparing the widths using optical microscope, scanning electron microscope, and atomic force microscope images (see Supplementary Figure S1 for details).

As shown in Fig. 1b, an applied charge current (*I*) across the injector contact creates a non-equilibrium spin density in graphene, which is then diffused in the channel and detected as a non-local voltage by a remote FM detector. However, in the local part of the channel, the applied *I* yields an electric field (E) in the graphene, and the direction of E significantly influences the non-equilibrium spin density under the injector contact for spin-injection and -extraction processes by inducing a drift force onto the spin-polarized carriers[30].

We started investigating the spin transport properties in the folded graphene channel with varying magnetic fields (B) along in-plane (y) and out-of-plane (z) directions. First, spin-valve measurements were performed by sweeping an in-plane magnetic field along the y-axis ($B_y$), which is the easy axis of the FM electrodes. The applied $B_y$ field switches the relative magnetization direction (M) of the detector from parallel to antiparallel orientation with respect to the injector, resulting in a switching in the non-local resistance ($R_{nl,sv}$ = $V_{nl,sv}$/I), as shown in Figure 1d, where $V_{nl,sv}$ is the measured voltage and *I* is the applied charge current at the injector FM contact. To ensure a reliable spin-transport signal and extract spin-transport parameters, we focused more on systematic Hanle spin-precession measurements by applying a perpendicular magnetic field ($B_z$) to the graphene plane. The $B_z$ induces spin-precession and dephases the diffusing spins in the channel, resulting in a decrease in the spin signal, as shown in Figure 1e. The spin signal magnitude is found to be strongly dependent on applied gate voltage ($V_g$) and applied injection bias current (I). The maximum amplitude of the measured Hanle spin-precession signal is of $\Delta V_{nl,Hanle}$ = 2.65 mV and $\Delta R_{nl,Hanle}$ = $\Delta V_{nl}$/I =88.5 Ω at *I* =-30 μA and $V_g$ =-10 V (Figure 1e).

As the Hanle spin-precession signal should be half of the corresponding spin-valve signal ($\Delta V_{nl,sv}=2\Delta V_{nl,Hanle}$)[15,18], the amplitude of the spin-valve signal is estimated to be $\Delta V_{nl,sv}$=5.3 mV and $\Delta R_{nl,sv}$ =177 Ω. The folded graphene channel (w ~ 0.3 μm) results in high contact resistance due to its small contact area, while maintaining a moderate value of the channel resistance $R_{ch}$. This leads to ideal spin impedance matching conditions for efficient spin injection and the creation of a large spin accumulation, $\Delta\mu$ > 20 meV. In comparison of the spin signal ($\Delta R_{nl,sv}$) in our folded-graphene non-local spin valve device, recent works show spin signal up to 130 Ω in graphene with $TiO_2$-seeded MgO[35], hBN[41] and $Al_2O_3$[42]. Usually, FM contacts with hBN, amorphous carbon, one-dimensional (1D) electrodes, MgO, SrO, and $TiO_2$ barrier give rise to a spin-valve signal in graphene in the range of 1 - 20 Ω[15,18,20,23,24,36–39,43–48], whereas transparent contacts without tunnel barriers[49] show a spin-valve signal of a few mΩ. Furthermore, spin-valve signals in 2D semiconductor black phosphorus with h-BN barrier show up to 50 Ω[50,51], and Si and GaAs show signals are in the range of 30 - 35 mΩ[52,53] at room temperature.

**Observation of large spin rectification effects:**

Such a large spin accumulation in our graphene channel can enable the observation of strong non-linear spin transport and spin rectification effect[26,30]. This is expected due to the nonlinear interaction between spin and charge currents, purely based on the energy dependence of the conductivity, without the need for any spin-orbit interaction or external magnetic fields. We performed electrical bias-dependent measurements of the spin signal to evaluate non-linear spin transport. As shown in Figure 1f, for positive bias current (+*I*), the spins are injected from the FM contacts into the graphene channel, resulting in the accumulation of the spins, whereas for negative bias current (-*I*), the spins are extracted by the FM from the graphene providing an opposite spin accumulation in the graphene channel. The injected or extracted spins create a non-equilibrium spin density in graphene that diffuses through the graphene channel, which is detected as a non-local voltage in Hanle spin-precession measurements. In a linear spin transport regime, the magnitude of the spin signal for both injection and extraction regimes should be linear and hence symmetric with respect to the bias current[4,54]. However, we observed a strong asymmetry in the magnitude of the Hanle spin signal for the spin-injection and -extraction processes (Figure 1g).

The measured Hanle spin-precession signals, along with the fitting at different applied bias currents (I), are systematically presented in Figures 2a and 2b. We can notice that reversing *I* render a sign change of the observed spin-signals as it creates opposite spin-polarization for spin-injection and -extraction processes at the junction. Figure 2c shows the spin-signal amplitudes ($\Delta V_{nl,Hanle}$ and $\Delta R_{nl,Hanle}=\Delta V_{nl,Hanle}/I$) of the observed voltage signals for positive and negative biases. A larger bias current gives rise to larger spin signals ($\Delta V_{nl,}$ $_{Hanle}$) because spin density at the injector electrode increases with applied bias current[18,55]. Interestingly, more than an order of magnitude bias-dependent asymmetry of the spin-signal amplitudes is observed, which is a hallmark of non-linear spin transport. For instance, we observe a larger $\Delta V_{nl}$=-2.29 mV ($\Delta R_{nl}$=57.4 Ω) with *I* =-40 μA and a smaller $\Delta V_{nl}$=0.189 mV ($\Delta R_{nl}$=4.73 Ω) with *I* =40 μA. While our study focuses on electron-doped conditions, the nonlinear spin transport mechanism should persist across all doping regimes, with signal amplitudes varying according to carrier concentration and mobility[31]. Similar spin amplification and rectification effects are observed for both positive and negative bias currents in another folded graphene spin-valve device, as shown in Supplementary Figure S2.

In our device, the large spin accumulation enables us to explore transport properties at energies away from the Fermi level, allowing each spin channel to experience a different conductivity. This phenomenon is valid if the conductivity (σ) is energy ($\epsilon$) dependent and can be parameterized by the coefficient, $\alpha = \frac{1}{\sigma}\frac{\partial\sigma}{\partial\epsilon}$, which is within the Drude model and using the Einstein relation is equivalent to $\alpha = \frac{\mu}{eD}$ (here, μ, D, and e are mobility, electron diffusion constant, and electron charge, respectively)[25,30].

The significant spin rectification effect observed in the spin signal can be understood as a manifestation of the nonlinear interaction between the charge current and the spin accumulation (see Figure 2d). The current charge is applied in the local injector contact, and the non-local detector FM contact is used to detect the spin accumulation (μ) on graphene in our measurement geometry. The spin injector contact creates a spin accumulation that, for low currents in the linear regime, diffuses away symmetrically from the injection point, on one side towards the nonlocal part of the circuit, and the other side towards the local part where a charge current is present, with a common decay length λ (light-colored plots in Figure 2e). The situation differs for the higher applied currents, where the electric field (E) in the local part of the circuit interacts with the injected spins (dark-colored plots in Figure 2e). In our device configuration, for *I* > 0, the E exerts a drift force on electrons carrying them away from the spin injection contact, resulting in a 'downstream' decay length $L_{-}$ > λ (plots with blue shades in Figure 2e)[25]. Thus, the spin accumulation profile extends up to a longer distance within the local part of the circuit, while reducing the spin density in the nonlocal part of the circuit and leading to a smaller nonlocal spin signal. The opposite situation is present for *I* < 0, where the E in the local part of the circuit exerts a drift force that carries electrons towards the spin injector, leading to an 'upstream' decay length within the local part of the circuit, $L_{+}$< λ (plots with dark-red shades in Figure 2e)[25]. In this case, the E aids in focusing the injected spin back towards the spin injector contact, resulting in an amplification of the spin accumulation in the nonlocal part of the circuit and a correspondingly larger nonlocal spin signal for the -*I* than the +*I* counterpart. We employ the formulation of nonlinear spin interaction[30] to our device to calculate the asymmetric spin accumulation profiles and the resulting nonlocal spin signal. The simulation results, shown in Figure 2f, show the higher-order effect on the nonlocal spin accumulation, with a satisfactory agreement with the experiment, both in terms of magnitude and modulation of the nonlocal spin resistance. Notably, our simulations indicate that we are creating spin accumulations at the injection point as large as $\Delta\mu_s$ of 20 meV at room temperature, which is much higher than previous reports[23,24]. The non-linear spin accumulation with $\pm I$ generates asymmetric $\Delta\mu_s$ in the folded-graphene channel, exhibiting the spin-diode effect that is more pronounced due to the very large spin accumulation. The detailed outline of modelling non-linear spin transport in folded graphene is provided in the Supplementary Information Section S1[25,30,57].

We find that the narrow-folded graphene structure is the key to the observation of a large signal amplitude and highly nonlinear spin-transport. The folded and narrower graphene channel offers lower channel resistance with a reduced contact area, leading to an enhanced spin impedance matching condition compared to a broader and unfolded graphene. Moreover, the high-quality FM contacts on the folded graphene spin-valve device lead to a near-perfect impedance matching condition, resulting in a larger signal amplitude in the highly nonlinear regime. For comparison, we also measured spin-transport properties in a multilayer (ML) unfolded graphene device (Supplementary Figure S3), where we observed a smaller spin signal and linear spin transport properties. A crucial distinction of folded bilayer graphene lies in its unique electronic structure, which fundamentally differs from conventional multilayer graphene systems. Rather than exhibiting the graphitic behavior typical of thick multilayer stacks, folded bilayer graphene maintains the intrinsic transport characteristics of individual bilayer units operating in a parallel configuration. This architecture preserves the microscopic spin transport physics characteristic of pristine bilayer graphene, offering a pathway to harness bilayer-specific phenomena while achieving enhanced spin signal amplitudes through parallel channel operation.

**Large gate tunability of spin signal:**

To understand the spin transport mechanism in such folded-bilayer graphene devices, we measured the Hanle spin-precession signals at various applied $V_g$ and found out the dependence of spin-signal ($R_{nl}$), spin polarization (P), spin-lifetime ($\tau_s$) and spin diffusion length ($\lambda_s$). Figure 3a presents the measured Hanle spin precession signals ($R_{nl,Hanle}$) along with the Hanle fitting[13,58] at various $V_g$ with $I$=−30 μA. The sign of the Hanle spin signal remains the same for the range of applied $V_g$, as expected for the non-local spin signal because of similar spin transport properties of electrons and holes in graphene[18,39]. However, the magnitude of the spin signal varies a lot with a maximum $V_g$=-10 V (Figure 3b), where the $\Delta R_{nl,Hanle}$=88.5 Ω and the expected spin-valve signal amplitude should be $\Delta R_{nl,sv}$=177 Ω. This large spin signal originates from efficient spin injection and detection in our hybrid Co/$TiO_2$/folded-bilayer graphene junctions due to near-ideal conductance matching conditions.

To estimate the spin-injection efficiency, we estimated the spin polarization of the contacts at various $V_g$ (as shown in Figure 3c) using equation (1)[59].

$$\Delta R_{NL} = \frac{P_i P_d R_{sq} \lambda_s}{W} e^{-L/\lambda_s} \qquad (1)$$

Where $P_i$, $P_d$, $\Delta R_{nl}$, $R_{sq}$, W, L, $\lambda_s$ are the injector's spin polarization, detector's spin polarization, spin-signal magnitude, graphene channel resistance, channel width (0.3 μm), channel length (1.55 μm), and spin diffusion length, respectively. We assume $P_i$ and $P_d$ to be equal and found a large spin polarization of P= 24.5% at $V_g$=-10 V, giving rise to efficient spin injection and detection in the device. It is to be mentioned that the metallic-oxide tunnel contacts possess pinholes and usually show less than 10% spin polarization[4,35,42,60–64]; whereas, we observed spin polarization (P) of 24.5% in Co/$TiO_2$/folded-graphene junctions. The enhancement of P in our device is most likely due to the narrower channel of the folded graphene.

Conductivity matching between the ferromagnetic tunnel contacts and spin transport channel is usually a major issue for achieving such an efficient spin injection and detection process. To correlate the variation of $\Delta R_{nl}$ and P with the applied $V_g$, we compared them with the $V_g$ dependence of channel resistance. We measured the gate dependence of folded-bilayer graphene resistance ($R_{ch}$) with a four-terminal measurement geometry (Figure 3d). The charge neutrality point is around $V_g$=-19.8 V and the corresponding channel resistance is 4.7 kΩ. The field effect mobility $\mu = \frac{L}{WC_g} . \frac{dG}{dV_g}$ is estimated to be 3500 $cm^2V^{-1}s^{-1}$; where $C_g$ and $dG/dV_g$ are gate capacitance per area ($1.15x10^{-8}$ $F.cm^{-2}$ for 300 nm $SiO_2$) and channel transconductance, respectively. Considering the spin-transport properties at various $V_g$, it is expected that for transparent and intermediate contacts (with injector's contact resistance, $R_i$<$R_{ch}$), $\Delta R_{nl}$ should show a dip at the charge neutrality point, whereas for tunneling contacts (with $R_i$>$R_{ch}$), $\Delta R_{nl}$ should be maximum[35]. In our device, the injector and detector contact resistances ($R_i$, $R_d$,~20 kΩ) are much higher than the graphene channel resistance ($R_{ch}$ ~ 4.7 kΩ). In the regime of $R_i$>$R_{ch}$, the spin-dependent chemical potential ($\Delta\mu$) is proportional to the $R_{ch}$ and given by, $\Delta\mu \propto R_{ch} \cdot I$ [35,65]. We observed the largest magnitude of spin signal close to the maximum of $R_{ch}$ near the charge neutrality point. Such gate dependence can be explained by the signal being limited by spin injection efficiency and satisfaction of conductivity matching rules at the Co/$TiO_2$/folded graphene injector contacts[35].

**Estimation of spin-transport parameters and conductivity matching condition:**

To get more insights into the spin-transport properties in the folded graphene channel, we extracted the bias-dependent spin-transport parameters from the Hanle fitting function. The bias-dependent spin-relaxation time ($\tau_s$), spin-diffusion constant ($D_s$) and spin-diffusion length ($\lambda_s = \sqrt{D_s\tau_s}$) are presented in Figure 4a. We found that the $\tau_s$ shows a similar trend to the spin-signal magnitude with bias currents (top panel in Figure 4a), as $\tau_s$ is larger in -*I* range in comparison to +*I*. The modulation of the estimated $\tau_s$ from the Hanle measurements could be due to the increase (or decrease) of the spin signal near the peak (low B field region) via the nonlinear interaction, leading to a narrowing or broadening of the peak. The $D_s$ and $\lambda_s$ at different bias currents are presented in the bottom panels of Figure 4a, respectively. Both of these parameters ($D_s$ and $\lambda_s$) show a similar trend with *I* and vary within the usual range observed in few-layer graphene[15,18,39,46,55,66]. Next, we extracted spin-transport parameters from the gate-dependent Hanle signals, as shown in Figure 4b. The $\tau_s$, $D_s$ and $\lambda_s$ vary within the usual range. We approximate the average parameters from all the gate-dependent Hanle measurements (total 9 measurements) and obtain average $\lambda_s$=2.05 μm, $\tau_s$=272 ps, $D_s$=0.015 $m^2s^{-1}$, comparable to the reported values in few-layer graphene[15,18,39,46,55,66–68].

To investigate the influence of the conductivity matching conditions on the observed large spin signal, we adapted a spin drift-diffusion model[65,69], considering the resistances of FM, folded-bilayer graphene and the tunnel contact. Figure 4c shows the estimated non-local spin-valve signal magnitude ($\Delta R_{nl, sv}$) as a function of injector FM/$TiO_2$/folded graphene interface resistance ($R_i$) for various gate voltages $V_g$. We fixed the detector FM/$TiO_2$/folded graphene interface resistance to 20 kΩ and used the respective spin-transport parameters for various $V_g$ as observed in the measurements. The calculations show that $\Delta R_{nl, sv}$ increases with increasing $R_i$ due to the enhancement of spin-injection efficiency and the reduction of the conductivity mismatch issue[39]. The dashed line in Figure 4c denotes the expected $\Delta R_{nl, sv}$ for the measured injector Co/$TiO_2$/folded graphene interface resistance. We found that $\Delta R_{nl,sv}$ is the highest for $V_g$=-10 V (near the charge neutrality point), followed by $\Delta R_{nl,sv}$ at $V_g$=40 V (electron-doped regime) and $\Delta R_{nl,sv}$ at $V_g$=-40 V (hole-doped regime) for any $R_i$. A similar trend in the magnitude of $\Delta R_{nl, Hanle}$ is observed for various $V_g$ in our measurements, as presented in Figure 3b. From these calculations, it appears that our spin-valve device operates close to the optimum spin injection efficiency limit.

**Discussions**

It is to be mentioned that, previously, diode behaviour was also observed using bilayer graphene due to spin drift effects[31], but with a limited spin accumulation with signal amplitude of around 0.5 Ω. However, a considerably larger spin accumulation of over 20 meV in folded bilayer graphene, with a significant signal of 177 Ω, provides a much-pronounced spin diode effect. Furthermore, thermoelectric effects[70] can lead to a background offset in non-local measurements; however, that does not affect the quantification of the measured spin-valve and Hanle signal magnitudes with magnetic field sweeps. Therefore, the observed large spin signal and spin-diode effect emerge from optimal impedance matching between ferromagnetic tunnel contacts and the folded graphene channel, coupled with nonlinear spin accumulation enhanced by carrier drift dynamics. However, a significant challenge also remains for the graphene spintronics field with the reproducible device performance, which requires precise optimization of multiple coupled parameters—spin polarization of ferromagnets, tunnel resistance and spin-polarized tunneling, graphene channel resistance, geometry, and carrier density—that must work synergistically to maintain optimal impedance matching for device applications. During the review process of our paper, the fabrication of rolled graphene has also been shown to be achievable, demonstrating significant chirality-induced spin selectivity[71]. At present, we do not have direct evidence of such spin-chiral effects, but they cannot be ruled out. The presence of synergetic effects demonstrates that folded bilayer graphene can simultaneously yield large spin signals and efficient rectification. Such spin-selective electron transport with active spintronic functionalities can be useful for future device architectures[4,22].

## Methods

### Device fabrication

The graphene is exfoliated from HOPG crystal and non-magnetic Cr/Au metallic contacts and ferromagnetic (FM) $TiO_2$(~1 nm)/Co(60 nm) contacts were fabricated using electron beam lithography and electron beam evaporation (as shown in Figure 1). The ferromagnetic (FM) $TiO_2$(~1 nm)/Co(60 nm) contacts were fabricated on the folded graphene channels using a two-step Ti deposition and oxidation process. The metal contacts were fabricated in both devices after the graphene had been rolled up. The contact resistance for the injector and detector contacts are about 20 kΩ and 24 kΩ, respectively. The folding of graphene occurred during the spin coating and lift-off processes. Specifically, 0.4 nm Ti was deposited two times and oxidized in situ with oxygen at 10 Torr for 10 min each time. Next, 60 nm Co is deposited by electron beam evaporation, followed by lift-off in acetone at 65 °C.

### Measurements

Spin transport measurements were performed at room temperature with a magnetic field up to 0.6 Tesla under vacuum conditions. Electronic measurements were carried out using current source Keithley 6221, nano-voltmeter Keithley 2182A, and dual-channel source meter Keithley 2612B.

## Declaration statements

## Data availability statements:

The datasets generated and/or analyzed during the current study are not publicly available due to ongoing further studies but are available from the corresponding author on reasonable request.

## Acknowledgment

The authors acknowledge financial support from European Commission Horizon Europe Graphene Flagship project “2D Heterostructure Non-volatile Spin Memory Technology” 2DSPIN-TECH (No. 101135853), 2D TECH VINNOVA competence center (No. 2019-00068), FLAG-ERA project 2DSOTECH (VR No. 2021-05925) and MagicTune, Wallenberg Initiative Materials Science for Sustainability (WISE) funded by the Knut and Alice Wallenberg Foundation, EU Graphene Flagship (Core 3, No. 881603), Swedish Research Council VR project grants (No. 2021-04821, No.

2018-07046), Graphene center, AoA Nano, Materials and Energy programs at Chalmers University of Technology. We acknowledge Horizon Europe Project "2D Heterostructure Non-volatile Spin Memory Technology" (2DSPIN-TECH) by UKRI, grant no. 10101734, University of Manchester, and ERC Project Twistrain, Flagera Multispin. We acknowledge the help of staff at the Quantum Device Physics and Nanofabrication laboratory in our department at Chalmers. Devices were fabricated at the Nanofabrication laboratory, Myfab, MC2, Chalmers.

## Author Contributions Statement

M.A.H and Z.K.K fabricated and measured the devices and equally contributed. B.Z. and P.R. supported in measurements and data analysis, I.V.M. analyzed the data and performed calculations, S.C., P.M., M.A.H and S.P.D wrote the manuscript with input from all co-authors. S.P.D supervised the research.

## Competing Interests

The Authors declare no Competing Financial or NonFinancial Interests.

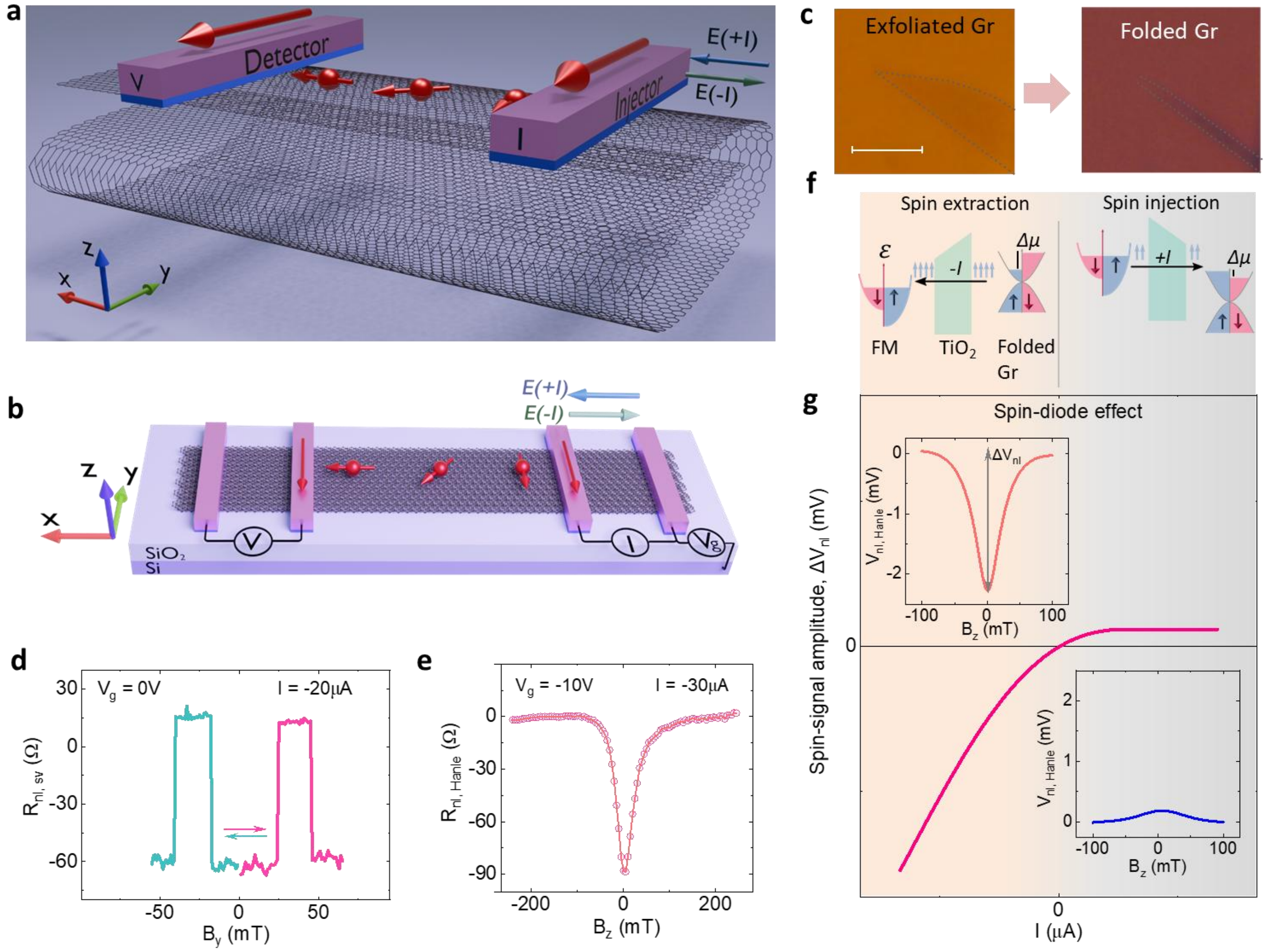

**Figure 1: Large spin signal with rectification effects in folded graphene at room temperature. (a)** Schematic of the spin-valve device with ferromagnetic (FM) contacts on top of a folded graphene. The directions of the associated electric field (E) to the applied bias currents (±*I*) are shown accordingly. **(b)** Graphical representation of the non-local measurement geometry of folded graphene spin-valve device on $SiO_2$/Si substrate. **(c)** Microscope picture of the graphene and folded graphene. The scale bar in the left image is 2 μm. The 1 μm wide graphene layer is folded 3-4 times to result in 0.3 μm wide graphene ribbon. **(d)** Non-local spin-valve signal in folded graphene with injection bias current of *I*=-20 μA at $V_g$=0 V (electron-doped regime) and 300 K. This spin-valve signal amplitude, $\Delta R_{nl,\,sv}=\Delta V_{nl}/I$ =75 Ω. The arrows indicate the directions of the magnetic field sweep. **(e)** The largest observed Hanle spin-precession signal $\Delta R_{nl,\,Hanle}$=88.5 Ω in the folded graphene device with *I* = -30 μA at $V_g$ = -10 V (electron-doped regime). **(f)** Schematics of spin-injection and extraction processes for the applied bias current (±*I*) in the folded graphene spin channel from ferromagnetic tunnel contact. The blue and pink regions in the energy band diagrams of the FM and folded graphene[40] indicate spin density of states for up(↑) and down (↓) spins, respectively. **(g)** Schematic representations of spin-signal amplitude as a function of applied bias current (I) showing spin-diode effect, where negative bias current (-*I*) renders higher spin signal amplitude in comparison to the spin signal with positive bias current

(+*I*). The insets show Hanle spin-precession signals for negative and positive bias currents to elucidate the spin-amplification and -rectification effects.

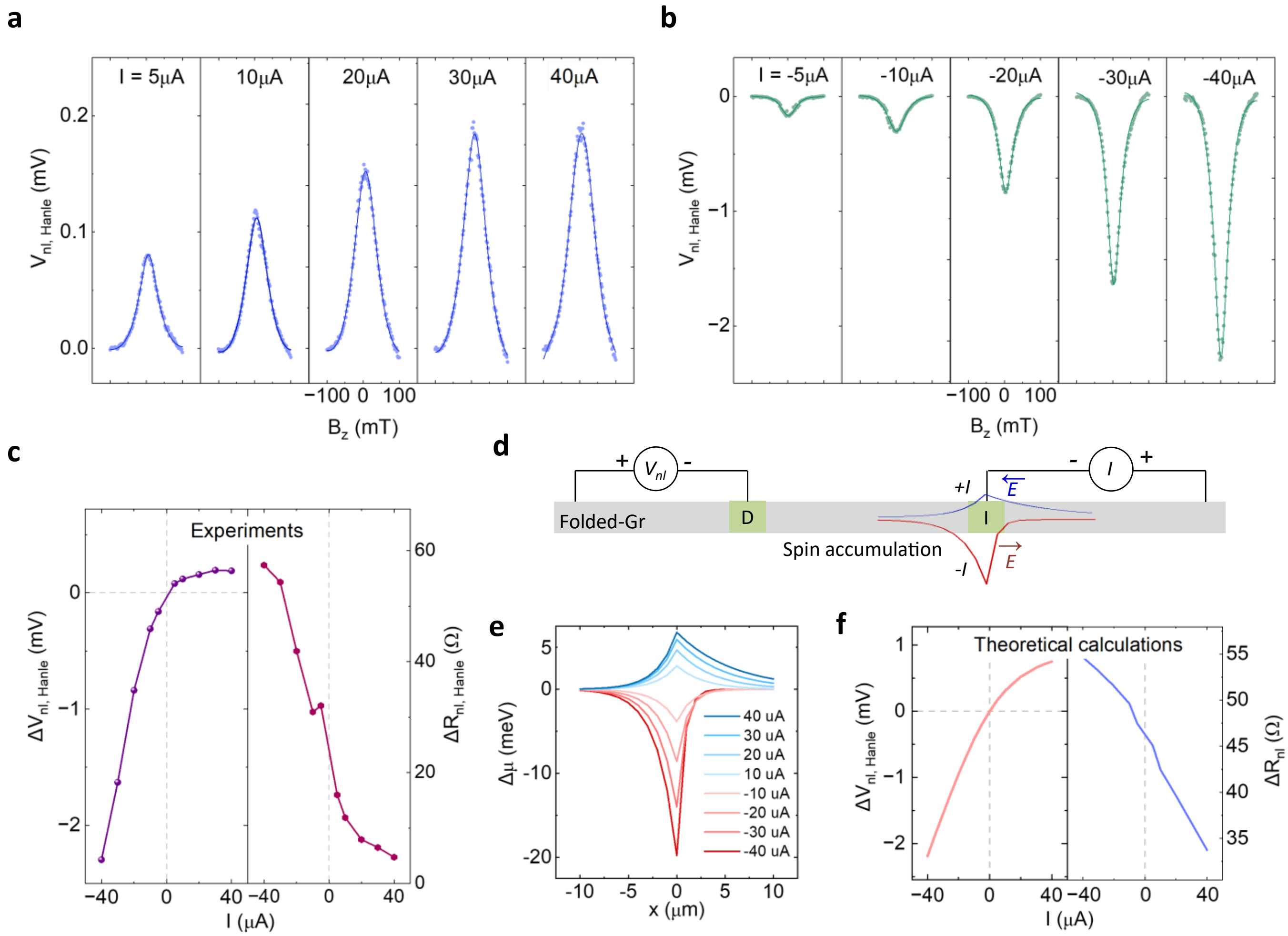

**Figure 2. Nonlinear spin transport and spin rectification effects. (a, b)** Hanle measurements (dots) for positive and negative bias currents *I* from ±5 μA to ±40 μA at $V_g$=40 V and 300 K along with the Hanle fitting (solid line). The measurements are conducted in electron-doped regime, where the carrier concentration[56] is about, $n_e$ = 4.30x$10^{12}$ $cm^{-2}$. A linear background is subtracted from the Hanle signals. **(c)** Bias dependence of Hanle signal amplitudes $\Delta V_{nl,\ Hanle}$ and corresponding $\Delta R_{nl,\ Hanle} = \Delta V_{nl,\ Hanle}/I$. **(d)** Non-local measurement geometry depicting bias current and associated electric-field (E) dependence of asymmetric spin-accumulation effect, where injector and detector FM contacts on graphene channel are denoted using "I" and "D". Blue and red lines highlights spin accumulation at the injector contacts, where negative bias current results in higher spin-accumulation with respect to the positive bias current. **(e)** Spin accumulation in the graphene channel for *+I* (plots with sky-blue shades) and *-I* (plots with dark-red shades) bias currents. In these calculations, we assume x = 0 is at the center of the injector electrode in our measurement geometry, as shown in Figure 2d. **(f)** Theoretical calculations of $\Delta V_{nl,\ Hanle}$ and $\Delta R_{nl,\ Hanle}$ with bias current considering non-linear spin transport.

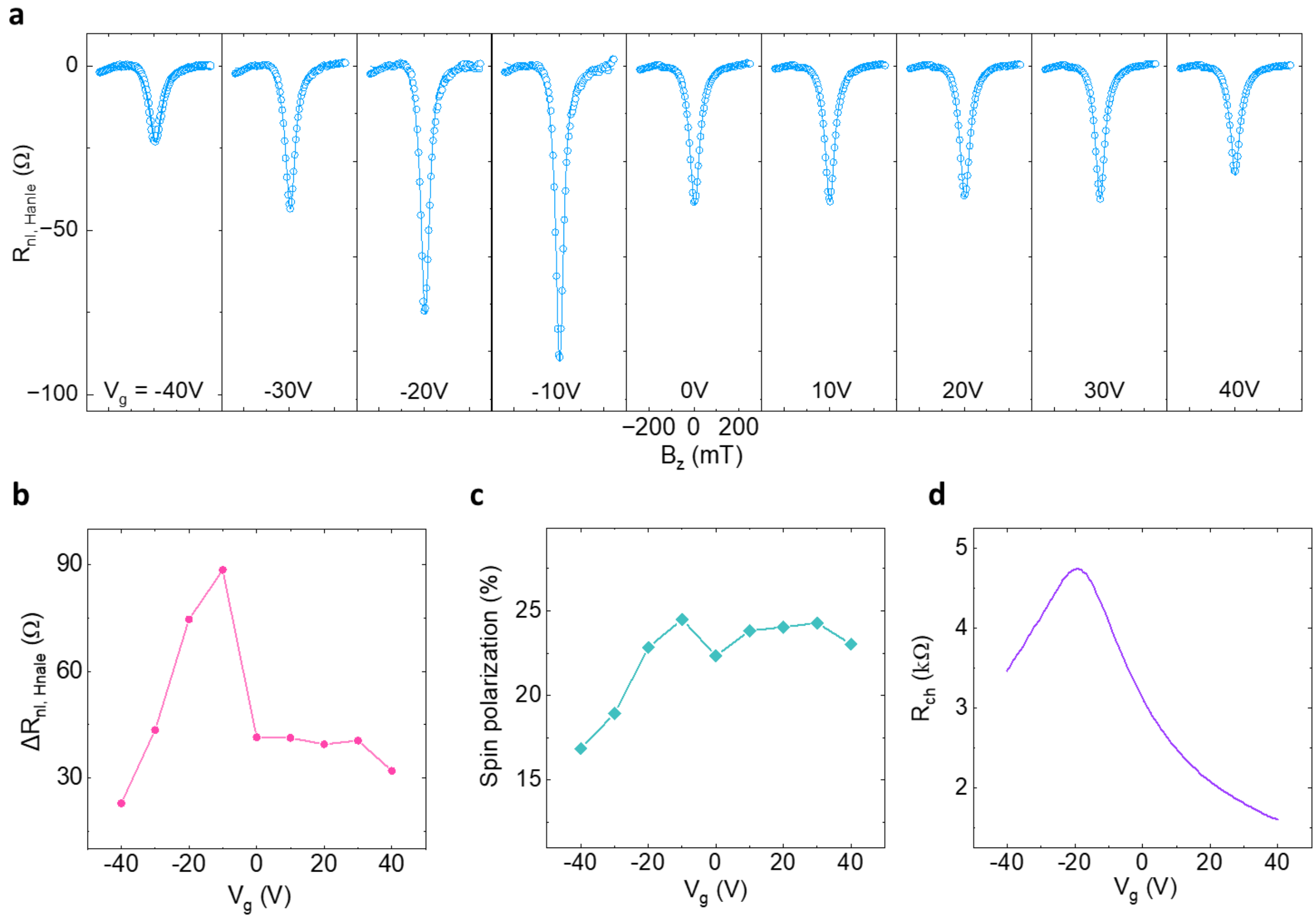

**Figure 3. Gate tunable spin signal. (a)** Gate dependence of the measured Hanle spin-precession signals ($R_{nl,Hanle}$ in circles) and associated Hanle fitting (solid line) with *I* =−30 μA at various $V_g$ in the range of -40 V to 40 V. A linear background is subtracted from the Hanle signal signals. **(b)** The gate-dependent magnitude of the Hanle spin-precession signal $\Delta R_{nl,Hanle}$. **(c)** Estimated spin polarization (P) in percentage at various $V_g$. (d) The gate-dependence of the graphene channel resistance ($R_{ch}$) with four-terminal measurement geometry.

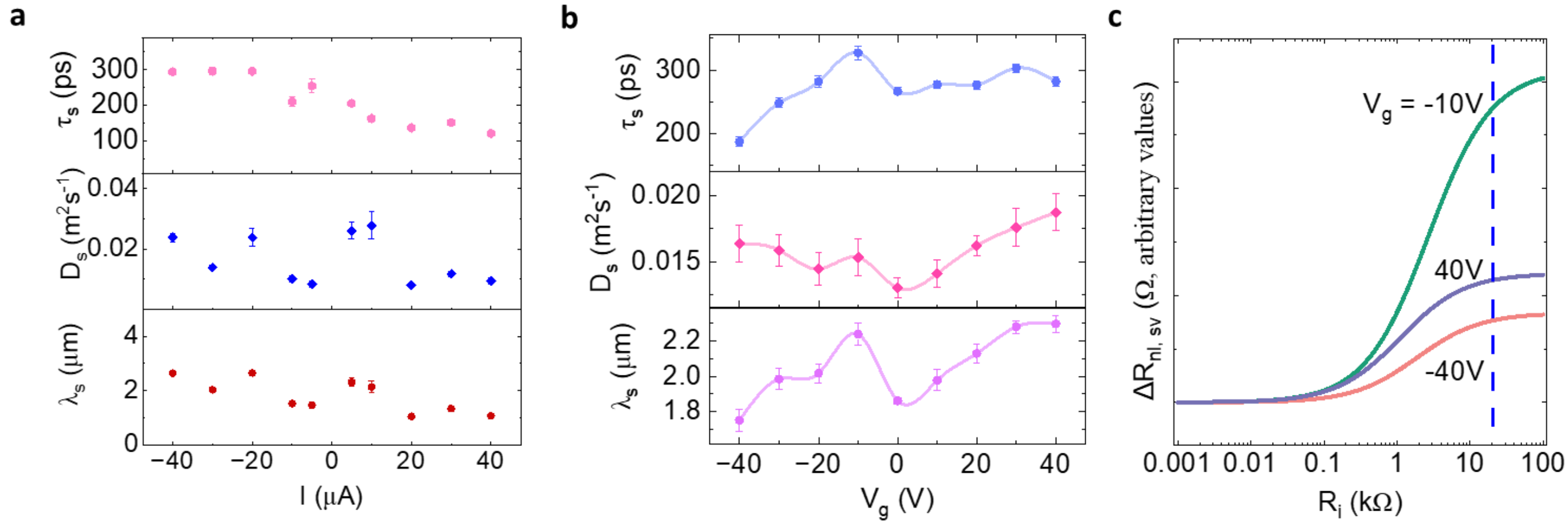

**Figure 4. Bias and gate dependence of spin-transport parameters. (a)** Bias-dependent spin relaxation time ($\tau_S$), spin diffusion constant ($D_s$), and spin diffusion length ($\lambda_s$) in folded graphene, which are obtained from the fitting of the Hanle spin-precession signals, presented in Figures 2a and 2b. **(b)** Gate voltage ($V_g$) dependence of the $\tau_S$, $D_s$, and $\lambda_s$ in folded graphene channel, which are obtained from the fitting of the Hanle spin-precession signals, presented in Figure 3a. **(c)** Estimation of spin signal magnitude ($\Delta R_{nl}$ in arbitrary Ω) as a function of the interface resistance between the injector FM and folded graphene channel at various $V_g$. The dashed line denotes the measured interface resistance (20 kΩ) of the injector's FM-folded graphene channel.